%% file: main-leakproof.tex
\PassOptionsToPackage{table,xcdraw}{xcolor}

\documentclass[acmtog,nonacm]{acmart}

\AtBeginDocument{%
  }

\usepackage{tabularx}
\usepackage{booktabs}
\usepackage{wrapfig}
\usepackage{algorithm}
\usepackage{algpseudocode}
\usepackage{subcaption}
\usepackage{graphicx}
\usepackage{tikz}
\usetikzlibrary{patterns}

\begin{document}

\title[Topology-Preserving Coupling of Compressible Fluids and Thin Deformables]{Topology-Preserving Coupling of \\
Compressible Fluids and Thin Deformables} 

\author{Jonathan Panuelos}
\email{jonathan.panuelos@mail.utoronto.ca}
\affiliation{%
  \institution{University of Toronto}
  \city{Toronto}
  \state{Ontario}
  \country{Canada}
}

\author{Eitan Grinspun}
\affiliation{%
  \institution{University of Toronto}
  \city{Toronto}
  \state{Ontario}
  \country{Canada}
}

\author{David Levin}
\affiliation{%
  \institution{University of Toronto}
  \city{Toronto}
  \state{Ontario}
  \country{Canada}
}

\renewcommand{\shortauthors}{Panuelos et al.}

\begin{abstract}
\input{sections/a-abstract}
\end{abstract}


\begin{teaserfigure}
    \centering
    \includegraphics[width=\textwidth]{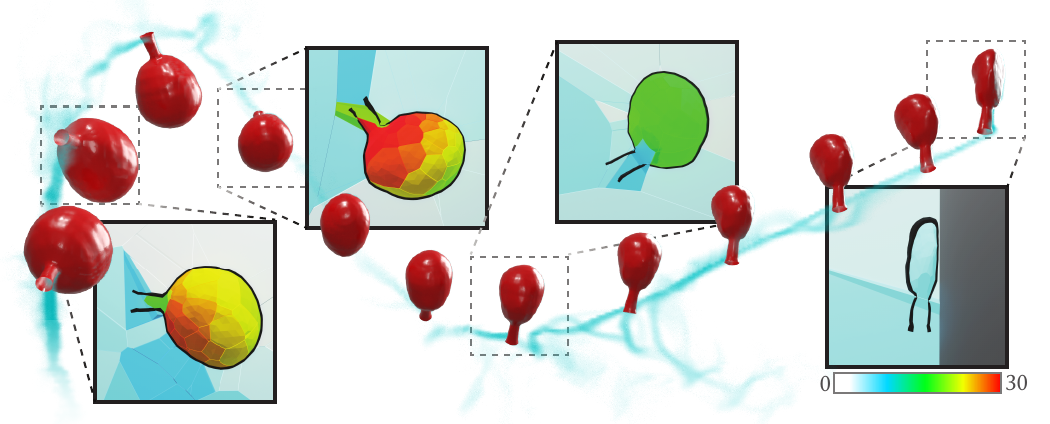}
    \caption{A balloon being inflated then released to be propelled via the energy transfer between elastic potential energy to air pressure to kinetic energy. Insets show pressure value at each Voronoi cell inside and around the balloon.}
    \label{fig:teaser}
\end{teaserfigure}

\maketitle

\input{sections/sec-introduction}
\input{sections/sec-background}
\input{sections/sec-equations_of_motion}
\input{sections/sec-discretization}

\input{sections/sec-numerical_flux}
\input{sections/sec-boundary_condition}
\input{sections/sec-examples}
\input{sections/sec-conclusion}

\bibliographystyle{ACM-Reference-Format}
\bibliography{bib}

\end{document}

%% file: sections/a-abstract.tex
We present a novel discretization of coupled compressible fluid and thin deformable structures that provides sufficient and necessary leakproofness by preserving the path connectedness of the fluid domain. Our method employs a constrained Voronoi-based spatial partitioning combined with Godunov-style finite-volume time integration. The fluid domain is discretized into cells that conform exactly to the fluid-solid interface, allowing boundary conditions to be sharply resolved exactly at the interface. This enables direct force exchange between the fluid and solid while ensuring that no fluid leaks through the solid, even when arbitrarily thin. We validate our approach on a series of challenging scenarios---including a balloon propelled by internal compressed air, a champagne cork ejecting after overcoming friction, and a supersonic asteroid---demonstrating bidirectional energy transfer between fluid and solid.

%% file: sections/sec-introduction.tex
\section{Introduction}
The interaction of compressible fluids with solids give rise to a range of varied and visually interesting phenomena.
From a balloon whizzing through the air propelled by its own elasticity, to car suspension absorbing road shocks, or a stomp rocket launching skyward, the coupling of compressible fluids and solids holds great potential for compelling simulations hitherto overlooked in computer graphics.
Beyond entertainment, these simulations have applications in areas such as modeling engine combustion chambers and capturing pressure gradients in biological systems like blood vessels.

While the general behaviour of fluids is governed by the Navier-Stokes equations, we focus our efforts specifically on the simulation of compressible inviscid fluids, which are given by the compressible Euler equations.
Compared to incompressible fluids commonly simulated in computer graphics where pressure is often taken to be the Lagrange multiplier for the incompressibility constraint, compressible fluid treat pressure as an extra state variable in a trio of conservation laws for mass, momentum, and energy.

In the process of attempting to resolve the coupling of compressible fluids to solids, prior work would destroy the topology of the boundary between the two phases.
They either insufficiently sample the boundary due to the difference in discretization between the solid and fluid phases, leading to leaking across thin boundaries, or sample these boundaries volumetrically thus adding thickness and potentially sealing flowable paths.
We point out that in scenarios with highly dynamic solid structures, this is problematic as it could lead to considerably different and incorrect behaviour.
Consider a self-propelled balloon for example, a leaky balloon might lose all its air before propelling itself, while an overly conservatively leakproof balloon may end up sealing its nozzle.

This motivates the need for a method that enforces \emph{necessary and sufficient leakproofness}, preserving the \emph{path connectivity} of the fluid domain. Fluid should be able to flow anywhere a continuous path exists through fluid, but never through solid.
Our key insight here is that to preserve all potential flowable paths, thus allowing flow if and only if this connectivity exists, the solid and fluid discretizations must agree on where the solid is.
That is, the solid geometry exactly must be a part of the fluid discretization.

We propose a novel fluid discretization that expands on prior Voronoi-based finite volume fluid discretizations \cite{springel2010pur} by introducing the solid geometry as faces in the fluid discretization.
Our new \emph{volume stitching algorithm} ensures that the fluid mesh conforms precisely to solid boundaries, while preserving the fluid domain's topology.
The result is an interface that is sufficiently and necessarily leakproof, ensuring that the discretized fluid domain respects the intended topology.
Furthermore, the interface is sharply resolved at the solid boundary, enabling accurate force exchange between solid and fluid. 

This partition can then be used for the explicit integration of the system, with fluid-fluid interactions being handled by the shared pairwise faces between any two particles, and the solid-fluid interactions are handled by any particles adjacent to the solid boundaries.
Information is accurately passed between the two phases, from solid to fluid via a Dirichlet boundary condition on velocity, and from fluid to solid via a pressure force.

We demonstrate the versatility of our method across a range of challenging scenarios---including a balloon propelled by escaping air, a champagne cork ejecting under pressure, and a supersonic asteroid generating Mach cones---examples involving strong bidirectional coupling and complex topology. These examples showcase the method’s ability to robustly handle thin structures, preserve fluid connectivity, and capture rich, visually compelling dynamics that were previously difficult to simulate within a unified framework.

%% file: sections/sec-background.tex
\section{Related Work}

\subsection{Compressible Fluids}

\subsubsection{Lagrangian Methods}
Much of the work in the simulation of compressible flow stem from the astrophysics community, which pioneered a Lagrangian description via Smoothed Particle Hydrodynamics (SPH) \cite{gingold1977smoothed, lucy1977numerical}. This involved a set of particles that track the fluid as it flows, with a kernel used for interpolation of physical quantities and their gradients onto the whole domain, allowing for low advection errors and automatic resolution adaptivity.

\subsubsection{Eulerian Methods}
In contrast, Eulerian methods partition space into \emph{static} cells, with advection being handled via some interpolation scheme \cite{stone1992zeus}. 
In these methods, the system is often solved via finite-volume Godunov schemes, with numerical fluxes being computed at each interface between cells. 
Cells may be structured or unstructured, with the choice of domain being problem dependent. 
While fully gaseous applications, such as those intended by \citet{stone1992zeus}, typically focus on structured grids, applications that require accurate boundary handling, such as those in engineering, demand the use of unstructured meshes.
\citet{mavriplis1997unified} presented a method for unstructured meshes consisting of  both triangular and quadrilateral faces that conforms to input geometry (in their intended application, plane wings). 
Our discretization similarly respects input solid geometry, but in a Lagrangian setting where source points can be in arbitrary locations in the fluid domain.
To achieve similar resolution adaptivity to Lagrangian methods, \citet{berger1989local} introduced adaptive mesh refinement (AMR) for locally reducing grid size.

\input{sections/img/fig-r-flatsheet}

\input{sections/img/fig-r-hourglass}

\subsubsection{Moving Unstructured Meshes}
Methods involving moving unstructured meshes attempt to achieve the advantages of both sides, with the high-accuracy flux and gradient computation of Eulerian schemes and accurate advection of Lagrangian. 
The domain is discretized using non-regular cells that move and deform according to the fluid motion. This grid motion is not necessarily required to follow the fluid velocity, allowing for an arbitrary Lagrangian-Eulerian (ALE) approach \cite{hirt1974arbitrary}, often used to avoid deteriorating mesh quality such as the presence of shards.

\citet{borgers2005lagrangian} introduced a Voronoi-based discretization, using it for incompressible flow, which was later adopted by \citet{serrano2005voronoi} and \citet{springel2010pur} for compressible problems.
The latter presented a volumetric approach to one-way solid coupling by explicitly including solid Voronoi points, but such volumetric approaches fail to properly resolve thin solids, as shown on Figure \ref{fig:r-flatsheet}, as well as potentially sealing small openings, shown on Figure \ref{fig:r-hourglass}.
Our work adopts their fluid discretization in the bulk flow, and presents a method for handling bidirectional solid coupling by modifying the Voronoi partition in the regions around the solid boundaries.

In contrast with the Voronoi representation of fluids, Voronoi partitions have also been used to represent solids in cell lattice methods, with \citet{hwang2021coupling} presenting a coupling scheme with weakly-compressible SPH fluids. We once again emphasize the limitations of such a volumetric representation of solids.

\subsubsection{Moving Meshless Methods}
With similar objectives, moving meshless methods have been developed, where source points partition space, but via fields of overlapping weights rather than strict mesh allocations \cite{lanson2008renormalized}. 
These differ from SPH in that they define a partition of unity over all space and still apply Godunov-like interface fluxes \cite{hopkins2015new}.

\subsubsection{Riemann Solvers}
As part of Godunov-type finite volume schemes, at each interface, the computation of a stable numerical flux is required. 
The first, and most diffusive, approximation is due to \citet{lax1954weak}, which is equivalent to taking an average over an entire cell's domain. \citet{kurganov2000new} reduces this domain of dependence to within the maximum signal velocity of waves propagating from the interface, reducing diffusivity.
Various other approximate solver have been proposed, such as the Roe linearization \cite{roe1981approximate}, HLLE \cite{harten1983upstream, einfeldt1988godunov}, HLLC \cite{toro1994restoration}, as well as exact solvers relying on Newton iteration \cite{toro2013riemann}, but we note that our method is agnostic to the choice of Riemann solver.

\subsubsection{In Computer Graphics}
Much of the effort in fluid simulation in computer graphics has largely been focused on incompressible flow.
SPH in particular, although having its roots in the highly compressible regime, has been adapted by the community for incompressible flow, via the application of different particle potentials \cite{desbrun1996smoothed} or via a pressure solve \cite{bender2016divergence}.
Some efforts have been done to soften the divergence constraint and allowing for controlled deviations from a defined rest density \cite{becker2007weakly, he2025semi}, but this work is limited to the weakly compressible regime.
The closest in this field to our work is that of \citet{cao2022efficient}, which simulate true compressible fluids in the context of supporting fluid shock-induced solid fracturing.
Due to their choice of an MPM discretization, their solids are inherently volumetric and as such are limited by the resolution of the MPM grid.
Our method, in comparison, conforms the fluid discretization around the solid, and are able to resolve even codimensional solids.

We achieve this surface-respecting discretization via a modification of the Voronoi diagram induced by the Lagrangian fluid particles.
The Voronoi diagram has been prior leveraged in computer graphics for representing fluid domains, albeit not compressible ones.
Incompressible flow was simulated by \citet{sin2009point}, computing the divergence free condition locally on each Voronoi cell.
\citet{brochu2010matching} added an explicit surface tracker that allows for computing more accurate surface forces such as surface tension. 
\citet{saye2011voronoi} similarly leveraged it for surface tracking for multiphase flows, aiding in simplifying topology changes.
Unlike these works, we consider every interface between each fluid particle, with physics being computed directly on the Voronoi diagram, rather than on a background Eulerian mesh. 
Voronoi diagrams have also been used to simulate foam bubbles, where each bubble is a radius-restricted Voronoi cell, with inter-bubble interfaces being represented by faces of the Voronoi diagram \cite{busaryev2012animating, qu2023power}. 

\input{sections/img/fig-r-hourglassnarrow}

\subsection{Solid-Fluid Coupling}

The domain of fluid-structure interaction (FSI) in mechanics literature is rich in methods for coupling fluid simulation to solids. Most notable among these are arbitrary Lagrangian-Eulerian (ALE) interfaces \cite{donea1982arbitrary}, the ghost fluid method \cite{fedkiw1999non}, and the immersed boundary (IB) method \cite{peskin1972flow}. 

\subsubsection{Arbitrary Lagrangian-Eulerian} methods discretize the fluid domain via a deformable mesh that attempts to match the mesh representation of the solid while retaining a fully static Euler representation far away from the solid \cite{donea1982arbitrary}.
The key difficulty in these methods is computing a valid deformation of the mesh without degrading mesh quality, thus limiting the method to small solid deflections of volumetric solids.
Recent advancements have been introduced to support incompressible fluids with codimensional solids with larger deformations via auxiliary coarse meshes, but deformations are still constrained and the method is limited to 2D and simpler geometries \cite{fernandes2019ale}.
These methods are thus not suitable to the significant deformations we present in our examples.

\subsubsection{Ghost Fluid Method} 
\citet{fedkiw1999non} introduced the ghost fluid method to provide sharp interface handling for Eulerian multiphase fluid simulations. Ghost values were introduced into the numerical stencil, modifying accessed data to enforce boundary conditions across interfaces, usually tracked via level sets.
The method was later extended from multiphase fluids into interfaces with Lagrangian solids \cite{fedkiw2002coupling}.
This work also became the preferred method for solid boundary handling in computer graphics literature \cite{bridson2015fluid}.

A similar method was adopted for boundary handling in SPH by filling solid domains with fictitious particles to avoid kernel deficiencies \cite{takeda1994numerical, randles1996smoothed}, which was also later adopted by the graphics community \cite{schechter2012ghost}. 
While this method is similar in spirit to the original ghost fluid, it loses sharp interface tracking due to the inherent smoothness introduced by SPH, as well as being necessarily volumetric. 

More recent work has introduced its use in arbitrary polyhedral finite volume discretizations \cite{vukvcevic2017implementation}. The key difficulty in these discretizations, shared from the original ghost fluid method, is the mismatch between the solid boundary and the fluid parcel boundary. We point out that the boundary condition becomes considerably simpler in our case because we match the fluid boundary to the solid boundary. We can simply reflect fluid particles across the face normal to enforce zero flux penetration across the solid faces.

\subsubsection{Immersed Boundary Method} embeds Lagrangian solid boundaries within a background Eulerian fluid grid, transferring forces and velocities by smoothing out the Lagrangian representation via delta functions \cite{peskin1972flow, peskin2002immersed}. 
Although initially developed for simulation of codimensional heart valves \cite{peskin1972flow}, no guarantees were made in terms of leakproofness.
In particular, because only the solid is treated as Lagrangian points, with the fluid being discretized on a static grid, the solid often passes by fluid control points.
In an incompressible fluid, as was the original implementation, this is not an issue as pressure and density remain constants throughout, but this offers fluid sidedness tracking challenges in compressible flows.
Additionally, the original formulation which blurs solid velocity across a finite thickness does not enforce exact velocity matching with the fluid, and can thus violate no-flux conditions.

Extensions were developed to allow for sharp enforcement of boundary conditions, both in incompressible \cite{mittal2008versatile} and compressible \cite{ghias2007sharp} flow.
These largely involve cut-cell and ghost cell approaches, effectively slicing the Eulerian grid to restrict it to one side of the domain.
We highlight the cut-cell methodology presented by \citet{ye1999accurate} (once again for incompressible flow), involving extending boundary grid cells to include empty cells missing their grid points cut off by solid boundaries.
These effectively modify the grid, warping the square grid into trapezoids wherever they are cut off by the solid boundary.
\citet{gretarsson2013fully} used a similar approach for compressible flow, and are able to handle thin solids, but recognized that their method may discard volumes that cannot be attached to adjoining grid cells, as shown on Figure \ref{fig:r-hourglassnarrow}.
Our stitching algorithm is conceptually similar to these approaches, reattaching orphaned fluid parcels to an existing fluid source point, and can be considered as a generalization of their method into unstructured meshes.
We demonstrate that our approach overcomes their limitations, and are able to resolve even subgrid fluid paths.

\subsection{Modified Voronoi Diagrams}

Visibility-constrained Voronoi diagrams have been extensively studied in computational geometry as generalizations of classical Voronoi tessellations to environments with occlusion or directional visibility restrictions. For instance, \citet{aurenhammer2014visibility} analyze diagrams where each site is restricted to a visibility wedge, resulting in Voronoi cells that may be non-convex, disconnected, and of quadratic combinatorial complexity. A broader treatment is provided in the monograph by \citet{okabe2000spatial}, which surveys visibility-aware and obstacle-constrained variants in applications ranging from robotics to spatial statistics. 

However, these constructions typically resolve endpoints of obstacles as Voronoi source points.
In contrast, our setting involves codimensional solid boundaries that \emph{lack} any Voronoi sources on their surface.
As a result, traditional visibility-constrained approaches cannot be directly applied. 
Using such methods adds thickness on solid vertices, potentially destroying the topology of the fluid domain.
Instead, we modify the Voronoi diagram induced solely by fluid particles by explicitly clipping it to the solid geometry and then reassigning orphaned cells via a path-connectivity-preserving stitching algorithm.
This constructively embeds solid interfaces into the fluid partition, enabling sharp boundary resolution and leakproof coupling—features not addressed by existing visibility-constrained methods.

\citet{tsin1996geodesic} does present a method for a Voronoi diagram where barriers only constrain visibility and do not induce their own sites, but their work is limited to rectilinear barriers in 2D.
The generalization to arbitrary barriers, as well as a further generalization to 3D, is nontrivial.

%% file: sections/img/fig-r-flatsheet.tex
\begin{figure}[t]
  \centering
  \includegraphics[width=\linewidth]{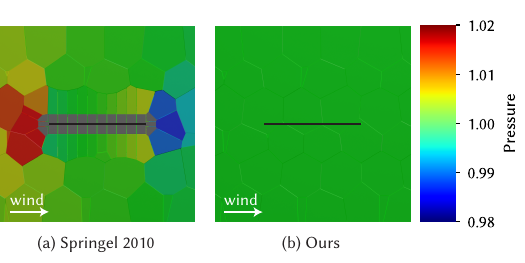}
  \caption{Simulation of an infinitessimally thin sheet immersed in an inviscid fluid moving rightwards. Volumetric methods such as that of \citet{springel2010pur} produce nonphysical pressure variation at the leading and trailing edges of the sheet. Solid cells are shown in grey, and the solid surface is shown as a black line.}
  \label{fig:r-flatsheet}
\end{figure}

%% file: sections/img/fig-r-hourglass.tex
\begin{figure}[t]
  \centering
  \includegraphics[width=\linewidth]{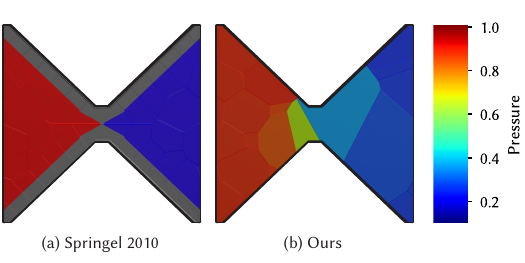}
  \caption{Cutaway view of an hourglass-shaped narrow opening initialized with a Sod shock tube, with the high-pressure region on the left and low-pressure region on the right. Approaches using solid particles, such as that by \citet{springel2010pur}, add additional thickness that closes the narrow opening, while our method resolves the solid \emph{at} the specified surface. Solid cells are shown in grey, the boundary is shown as a black outline.}
  \label{fig:r-hourglass}
\end{figure}

%% file: sections/img/fig-r-hourglassnarrow.tex
\begin{figure}[t]
  \centering
  \includegraphics[width=\linewidth]{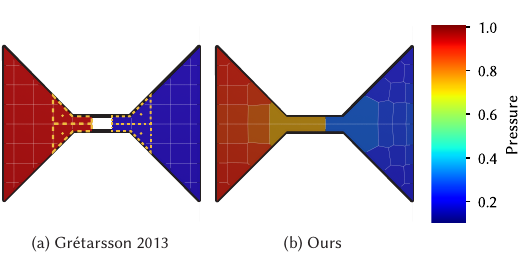}
  \caption{2D Sod shock tube through a narrow opening, with the boundary shown in black. The mass lumping of \citet{gretarsson2013fully} around the tube is shown in dotted yellow, discarding cells not adjacent to cell centers (yellow circles).}
  \label{fig:r-hourglassnarrow}
\end{figure}

%% file: sections/sec-equations_of_motion.tex
\section{Equations of Motion}

\input{sections/img/fig-methods-voronoi}

We aim to solve the fluid flow as described by the compressible Euler equations, which are a hyperbolic conservation law
\begin{align}
    \frac{\partial}{\partial t}\mathbf{U}(\mathbf{x},t) + \nabla\cdot\mathbf{F}(\mathbf{x},t) = \mathbf{0} \ , \label{eq:conservation_law}
\end{align}
where $\mathbf{U}$ represents the vector of conserved quantities
\begin{align}
    \mathbf{U} &= \begin{bmatrix}
    \rho & \rho u_x & \rho u_y & \rho u_z & \rho e_T
    \end{bmatrix}^\intercal  \ , \label{eq:state_vector}
\end{align}
consisting of mass density $\rho$, momentum density $\begin{bmatrix}\rho u_x & \rho u_y & \rho y_z\end{bmatrix}^\intercal$, and total energy density $\rho e_T$, and
$\mathbf{F}$ is a nonlinear convective flux
\begin{align}
    \mathbf{F} &= \begin{bmatrix}
        \rho u_n \\
        \rho u_x u_n + P n_x \\
        \rho u_y u_n + P n_y \\
        \rho u_z u_n + P n_z \\
        \rho u_n \left(e_T + \frac{P}{\rho}\right)
    \end{bmatrix}
\end{align}
governing the flow of these conserved variables across some unit normal vector $\mathbf{n}=\begin{bmatrix}n_x & n_y & n_z\end{bmatrix}^\intercal$ arising from
the velocity $\mathbf{u}=\begin{bmatrix}u_x & u_y & u_z\end{bmatrix}^\intercal$
and normal velocity
$u_n=\mathbf{u}\cdot\mathbf{n}$. The specific total energy
$e_T=e+\frac{1}{2}||\mathbf{u}||^2$ is the sum of internal energy $e$ and kinetic energy $\frac{1}{2}||\mathbf{u}||^2$.

Observe the explicit presence of pressure $P$, rather than the usual Lagrange multiplier treatment used in incompressible simulation.

This system is underdetermined and requires an equation of state to complete it.
We use the ideal gas law $P=(\gamma - 1)\rho e$, but note that the method presented is agnostic of the chosen state equation. 
We take the adiabatic index $\gamma$ to be $1.4$, representing ideal diatomic gases.

Via Green's theorem, we can rewrite the conservation law as
\begin{align}
    \frac{\partial}{\partial t}\int_\Omega \mathbf{U}(\mathbf{x},t)\ dV + \int_{\partial\Omega}\mathbf{F}(\mathbf{x},t)\cdot\mathbf{n}\ ds = \mathbf{0}, \label{eq:weak_form}
\end{align}
which holds for any choice of fluid parcel $\Omega$ in space. This weak form guides our choice of discretization and naturally motivates the use of finite volume partitions, where numerical fluxes are computed at cell interfaces to enforce conservation locally.

%% file: sections/img/fig-methods-voronoi.tex
\begin{figure}[t]
  \centering
  \begin{subfigure}[t]{0.49\linewidth}
    \centering
      \includegraphics[width=\linewidth]{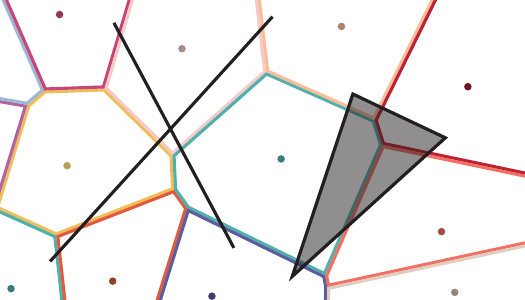}
    \caption{Initial Voronoi diagram (coloured outlines), clipped by solid constraints (black outlines, black shaded).}
    \label{fig:meshing-1}
  \end{subfigure}
  \begin{subfigure}[t]{0.49\linewidth}
    \centering
      \includegraphics[width=\linewidth]{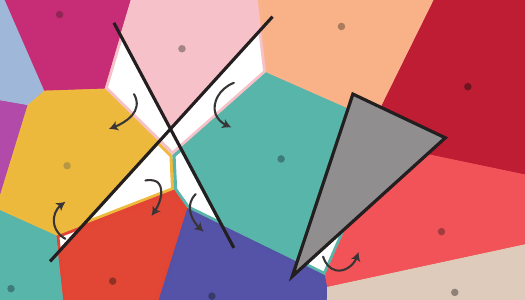}
    \caption{Cells containing a source point are assigned (filled), while the rest are registered as ``orphaned'' (white). Orphan cells will inherit the valid neighbours with the largest face area (grey arrow).}
    \label{fig:meshing-2}
  \end{subfigure}
  \begin{subfigure}[t]{0.49\linewidth}
    \centering
      \includegraphics[width=\linewidth]{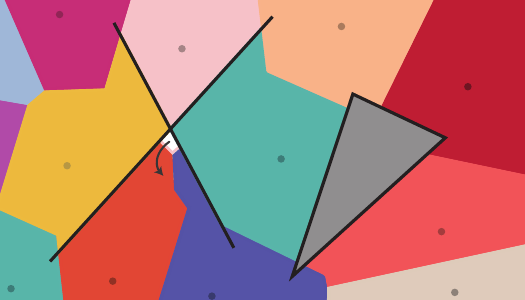}
    \caption{New connectivity reveals more orphaned cells that can be inherited.}
    \label{fig:meshing-3}
  \end{subfigure}
  \begin{subfigure}[t]{0.49\linewidth}
    \centering
      \includegraphics[width=\linewidth]{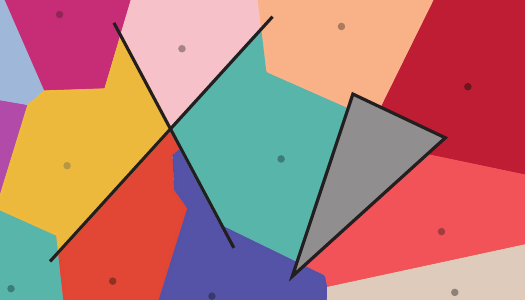}
    \caption{Final modified Voronoi diagram.}
    \label{fig:meshing-4}
  \end{subfigure}
  
  \caption{Stitching orphaned cells back to valid cells from the clipped Voronoi based on path-connectivity constraints.} \label{fig:methods-voronoi}
  \Description{Stitching orphaned cells back to valid cells from the clipped Voronoi based on path-connectivity constraints.}
\end{figure}

%% file: sections/sec-discretization.tex
\section{Topology-Preserving Discretization}

\input{sections/img/fig-methods-local_voronoi}
We seek a discretization that resolves boundary conditions imposed by arbitrarily shaped solids, including thin shells and other codimensional interfaces. This rules out fixed-resolution methods and volumetric solid representations, which require excessive refinement to capture thin structures. We also require support for large solid deformations—such as those of a deflating balloon—without compromising the fluid discretization, excluding static meshes and deformable mesh methods with fixed connectivity (e.g., ALE), which are typically limited to low-deformation volumetric solids.

These constraints naturally point to Lagrangian fluid representations. Among these, we prioritize compatibility with finite volume methods to sharply resolve solid interfaces. We adopt the Voronoi-based discretizations of \citet{borgers2005lagrangian} and \citet{springel2010pur}, in which Lagrangian particles track the fluid motion and induce a spatial partition, assigning each particle the region closest to it in space.

Applying this discretization to the weak form in Equation \ref{eq:weak_form} gives the fluid state update for some particle $i$ with nearest-neighbour particles $j$,
\begin{align}
    \frac{\partial}{\partial t}\mathbf{U}_i + \sum_j \frac{A_{ij}}{V_i}\hat{\mathbf{F}}_{ij}\cdot\mathbf{n}_{ij} &= \mathbf{0},\label{eq:godunov}
\end{align}
where $\mathbf{U}_i$ is the fluid state of particle $i$, $\hat{\mathbf{F}}_{ij}$ is a numerical flux computed between $i$ and $j$, $A_{ij}$ is the area of their interface, $n_{ij}$ is the outwards pointing normal of said interface, and $V_i$ is the volume associated with $i$. The geometry of this interface between particles $i$ and $j$ is shown in the inset above. 

This finite-volume equation where a numerical flux is evaluated at each interface is known as the Godunov form.

\subsection{Challenges to Leakproofing}

Significant care needs to be taken near boundaries, given that in our Godunov-type discretization (Equation \ref{eq:godunov}), forces are applied only along faces of the Voronoi partition. 
In order to properly couple any boundary condition with the fluid particles, we must modify the Voronoi diagram to include the boundaries as edges in the diagram.
Additionally, given a particle on one side of the barrier, its given domain should not extend past a barrier unless there is a contiguous path through the fluid to that side. In other words, a particle's domain must represent a single contiguous fluid.

\citet{springel2010pur} applied the solid boundaries as explicit particles that induce their own Voronoi cells.
This has the disadvantage of being a volumetric representation of the solid, and thus cannot accurately capture arbitrarily thin structures.

Figure \ref{fig:r-flatsheet}
demonstrates that the ``thickening'' approach is poorly suited to for immersed thin structures.
We simulate a flat sheet immersed in a tangentially flowing fluid.
Because the inviscid fluid flow is always parallel to the thin solid surface, the solid should not induce any change in the fluid state.
Unfortunately, the volumetric approach requires a finite thickness inducing a pressure build-up on the leading edge of the sheet and a low-pressure wake on the trailing edge.

Furthermore, as depicted in Figure \ref{fig:r-hourglass}, the ``thickening'' approach is
poorly sited for flow through small orifices, as it is prone to numerical sealing of small holes, such as those found in funnels, medicine droppers, jet sprays, or balloon nozzles.
Our method resolves orifices, no matter how small, without introducing additional particles to represent the solid surface.

A similar issue arises from the work of \citet{gretarsson2013fully} for a different reason.
Their discretization utilizes a static grid, and performs mass lumping to reallocate partial cells to neighbouring grid degrees of freedom. 
This has the issue of potentially discarding partial cells, as shown on Figure \ref{fig:r-hourglassnarrow}.

Therefore, we seek a discretization that exactly recovers the connectivity produced by the solid boundaries, is leakproof where and only where required, and allows for fluid flow wherever a valid path exists. 

\input{sections/algo-meshing}

\subsection{Clipped Voronoi Stitching Algorithm}

We develop a stitching algorithm that augments the set of interfaces from the Voronoi diagram with the faces of the solid boundary, reassigning patches of space to the appropriate fluid particle in a path-connected manner, so that fluid may continue to flow wherever a valid path exists around the solid.

Our method is initialized by constructing the standard Voronoi diagram induced by the fluid particles over all of space.
We then include all solid faces into this structure by clipping all the cells with these faces.
Any Voronoi faces inside volumetric solids will be removed, while codimensional solid faces will be added as new faces in the diagram.
Each fluid face is then tagged with the label of the bordering fluid cells, constructing a neighbourhood graph where each edge is a valid fluid-only path through space.

Around these solid faces, a number of cells will become ``orphaned'', \emph{i.e.} they will no longer contain a source point within the cell.
We tag these cells, then for each orphaned cell, we will assign it to its non-orphaned neighbour with the largest interface area, as shown on Figure \ref{fig:methods-voronoi}.
We choose to attach by largest interface area due to our choice of finite-volume as the integration method.
Since fluxes are computed pairwise at each interface weighted by interface area, orphaned cells ought to be most influenced by whichever neighbour it shares the largest face with.
This procedure is done iteratively, as some orphaned cells may be landlocked by other orphaned cells until said cells are assigned.

The completed algorithm is presented on Algorithm \ref{alg:meshing}. Our orphan-cell reassignment is conceptually similar to breadth-first region growing or flood fill, but adapted to prioritize interface area and operate on clipped Voronoi cells, preserving finite-volume structure and fluid path connectivity under solid constraints.

\subsection{Necessary and Sufficient Leakproofing}

\input{sections/img/fig-methods-flows}

It is straightforward to show that our method preserves the connectivity of the fluid domain as constrained by the input solid boundaries. 

First, solid boundaries are included as extra faces on top of the original Voronoi diagram.
Second, a fluid face is never deleted unless it is fully inside a volumetric solid.

Third, a fluid face split by a solid still persists and therefore still produces fluxes.
Such a face either belongs to a valid or an orphaned cell.
In the former case, it allocates its flux to its original valid cell.
In the latter case, we know that the orphaned cell will be connected to a valid cell through a series of shared fluid faces.
Therefore, any flux this face experiences represents some fluid that is able to make its way to the valid cell's particle via a path entirely within the fluid.
As such, the face may allocate its flux to that valid cell, representing a physically viable flow.

Conversely, the presence of any solid face inhibits fluid flow by disconnecting the connectivity of either side of the face.

Thus, our partition explicitly preserves the connectivity of the fluid domain induced by the solid boundaries.

Because our partition preserves this connectivity, we produce a method that is sufficiently leakproof, but also leakproof only when necessary, allowing for fluid flux where the original solid boundary conditions allow for.

%% file: sections/img/fig-methods-local_voronoi.tex
\begin{wrapfigure}{r}{1.2in}
  \centering
  \vspace{-0.5em} 
  \hspace{-2em}
  \includegraphics[width=\linewidth]{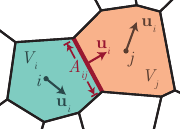}
  \label{fig:methods-local_voronoi}
  \vspace{-1.5em}
\end{wrapfigure}

%% file: sections/algo-meshing.tex
\begin{algorithm}[t]
\caption{Clipped Voronoi Stitching}
\label{alg:meshing}
\begin{algorithmic}[1]
\State Compute initial Voronoi tessellation
\State Clip Voronoi cells by solid boundaries
\For{each cell}
    \If{cell does not contain its generating point}
        \State Label cell as \texttt{orphaned}
    \EndIf
\EndFor
\While{any cell is orphaned}
    \For{each orphaned cell}
        \For{each neighboring non-orphaned cell}
            \State Compute interface area between cells
        \EndFor
        \State Merge orphaned cell with neighboring non-orphaned cell with largest interface area
        \State Update orphaned cell label
    \EndFor
\EndWhile
\end{algorithmic}
\end{algorithm}

%% file: sections/img/fig-methods-flows.tex
\begin{figure}[t]
  \centering
  \begin{subfigure}[t]{0.49\linewidth}
    \centering
      \resizebox{\linewidth}{!}{
      \input{sections/tikz/tikz-2dflows-1}
      }
    \caption{Voronoi mesh}
    \label{fig:methods-voronoi-flows-1}
  \end{subfigure}
  \hfill
  \begin{subfigure}[t]{0.49\linewidth}
    \centering
      \resizebox{\linewidth}{!}{
      \input{sections/tikz/tikz-2dflows-2}
      }
    \caption{Flux domains of dependence at each interface}
    \label{fig:methods-voronoi-flows-2}
  \end{subfigure}
  \caption{A Voronoi diagram at some timestep (left), and the domains of dependence of the waves propagating at each interface (right). Voronoi mesh is represented in green, with source points shown in red. Domains of dependence are shown in light blue.}
  \label{fig:methods-voronoi-flows}
\end{figure}
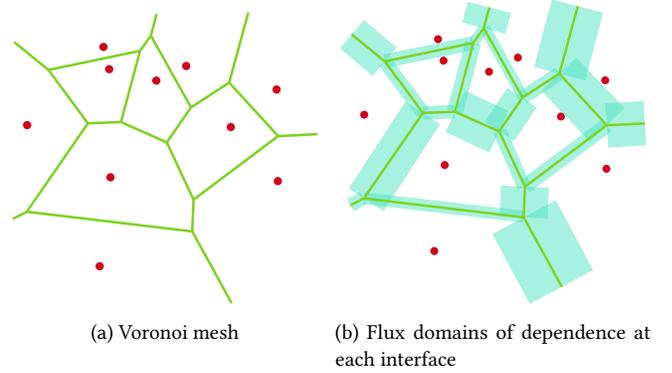

%% file: sections/tikz/tikz-2dflows-1.tex
\tikzset{every picture/.style={line width=0.75pt}} 

\begin{tikzpicture}[x=0.75pt,y=0.75pt,yscale=-1,xscale=1]

\draw  [draw opacity=0][fill={rgb, 255:red, 208; green, 2; blue, 27 }  ,fill opacity=1 ] (30,184.5) .. controls (30,180.91) and (32.91,178) .. (36.5,178) .. controls (40.09,178) and (43,180.91) .. (43,184.5) .. controls (43,188.09) and (40.09,191) .. (36.5,191) .. controls (32.91,191) and (30,188.09) .. (30,184.5) -- cycle ;
\draw [color={rgb, 255:red, 126; green, 211; blue, 33 }  ,draw opacity=1 ][line width=3]    (14,332) -- (36,320.67) ;
\draw [color={rgb, 255:red, 126; green, 211; blue, 33 }  ,draw opacity=1 ][line width=3]    (36,320.67) -- (296,351.67) ;
\draw [color={rgb, 255:red, 126; green, 211; blue, 33 }  ,draw opacity=1 ][line width=3]    (358,464.67) -- (296,351.67) ;
\draw [color={rgb, 255:red, 126; green, 211; blue, 33 }  ,draw opacity=1 ][line width=3]    (298,301.67) -- (296,351.67) ;
\draw [color={rgb, 255:red, 126; green, 211; blue, 33 }  ,draw opacity=1 ][line width=3]    (431,201.67) -- (298,301.67) ;
\draw [color={rgb, 255:red, 126; green, 211; blue, 33 }  ,draw opacity=1 ][line width=3]    (298,301.67) -- (256,211.67) ;
\draw [color={rgb, 255:red, 126; green, 211; blue, 33 }  ,draw opacity=1 ][line width=3]    (36,320.67) -- (132,181.67) ;
\draw [color={rgb, 255:red, 126; green, 211; blue, 33 }  ,draw opacity=1 ][line width=3]    (132,181.67) -- (184,179.67) ;
\draw [color={rgb, 255:red, 126; green, 211; blue, 33 }  ,draw opacity=1 ][line width=3]    (184,179.67) -- (256,211.67) ;
\draw [color={rgb, 255:red, 126; green, 211; blue, 33 }  ,draw opacity=1 ][line width=3]    (294,156.67) -- (256,211.67) ;
\draw [color={rgb, 255:red, 126; green, 211; blue, 33 }  ,draw opacity=1 ][line width=3]    (493,198.67) -- (431,201.67) ;
\draw [color={rgb, 255:red, 126; green, 211; blue, 33 }  ,draw opacity=1 ][line width=3]    (354,117.67) -- (431,201.67) ;
\draw [color={rgb, 255:red, 126; green, 211; blue, 33 }  ,draw opacity=1 ][line width=3]    (294,156.67) -- (354,117.67) ;
\draw [color={rgb, 255:red, 126; green, 211; blue, 33 }  ,draw opacity=1 ][line width=3]    (214,67.67) -- (184,179.67) ;
\draw [color={rgb, 255:red, 126; green, 211; blue, 33 }  ,draw opacity=1 ][line width=3]    (70,97.67) -- (132,181.67) ;
\draw [color={rgb, 255:red, 126; green, 211; blue, 33 }  ,draw opacity=1 ][line width=3]    (70,97.67) -- (214,67.67) ;
\draw [color={rgb, 255:red, 126; green, 211; blue, 33 }  ,draw opacity=1 ][line width=3]    (70,97.67) -- (16,53.67) ;
\draw [color={rgb, 255:red, 126; green, 211; blue, 33 }  ,draw opacity=1 ][line width=3]    (214,67.67) -- (231,43.67) ;
\draw [color={rgb, 255:red, 126; green, 211; blue, 33 }  ,draw opacity=1 ][line width=3]    (231,43.67) -- (294,156.67) ;
\draw [color={rgb, 255:red, 126; green, 211; blue, 33 }  ,draw opacity=1 ][line width=3]    (231,43.67) -- (240,9.67) ;
\draw [color={rgb, 255:red, 126; green, 211; blue, 33 }  ,draw opacity=1 ][line width=3]    (354,117.67) -- (384,7.67) ;
\draw  [draw opacity=0][fill={rgb, 255:red, 208; green, 2; blue, 27 }  ,fill opacity=1 ] (161,266.5) .. controls (161,262.91) and (163.91,260) .. (167.5,260) .. controls (171.09,260) and (174,262.91) .. (174,266.5) .. controls (174,270.09) and (171.09,273) .. (167.5,273) .. controls (163.91,273) and (161,270.09) .. (161,266.5) -- cycle ;
\draw  [draw opacity=0][fill={rgb, 255:red, 208; green, 2; blue, 27 }  ,fill opacity=1 ] (144,406.5) .. controls (144,402.91) and (146.91,400) .. (150.5,400) .. controls (154.09,400) and (157,402.91) .. (157,406.5) .. controls (157,410.09) and (154.09,413) .. (150.5,413) .. controls (146.91,413) and (144,410.09) .. (144,406.5) -- cycle ;
\draw  [draw opacity=0][fill={rgb, 255:red, 208; green, 2; blue, 27 }  ,fill opacity=1 ] (159,96.5) .. controls (159,92.91) and (161.91,90) .. (165.5,90) .. controls (169.09,90) and (172,92.91) .. (172,96.5) .. controls (172,100.09) and (169.09,103) .. (165.5,103) .. controls (161.91,103) and (159,100.09) .. (159,96.5) -- cycle ;
\draw  [draw opacity=0][fill={rgb, 255:red, 208; green, 2; blue, 27 }  ,fill opacity=1 ] (150,61.5) .. controls (150,57.91) and (152.91,55) .. (156.5,55) .. controls (160.09,55) and (163,57.91) .. (163,61.5) .. controls (163,65.09) and (160.09,68) .. (156.5,68) .. controls (152.91,68) and (150,65.09) .. (150,61.5) -- cycle ;
\draw  [draw opacity=0][fill={rgb, 255:red, 208; green, 2; blue, 27 }  ,fill opacity=1 ] (233,114.5) .. controls (233,110.91) and (235.91,108) .. (239.5,108) .. controls (243.09,108) and (246,110.91) .. (246,114.5) .. controls (246,118.09) and (243.09,121) .. (239.5,121) .. controls (235.91,121) and (233,118.09) .. (233,114.5) -- cycle ;
\draw  [draw opacity=0][fill={rgb, 255:red, 208; green, 2; blue, 27 }  ,fill opacity=1 ] (280,91.5) .. controls (280,87.91) and (282.91,85) .. (286.5,85) .. controls (290.09,85) and (293,87.91) .. (293,91.5) .. controls (293,95.09) and (290.09,98) .. (286.5,98) .. controls (282.91,98) and (280,95.09) .. (280,91.5) -- cycle ;
\draw  [draw opacity=0][fill={rgb, 255:red, 208; green, 2; blue, 27 }  ,fill opacity=1 ] (350,187.5) .. controls (350,183.91) and (352.91,181) .. (356.5,181) .. controls (360.09,181) and (363,183.91) .. (363,187.5) .. controls (363,191.09) and (360.09,194) .. (356.5,194) .. controls (352.91,194) and (350,191.09) .. (350,187.5) -- cycle ;
\draw  [draw opacity=0][fill={rgb, 255:red, 208; green, 2; blue, 27 }  ,fill opacity=1 ] (424,272.5) .. controls (424,268.91) and (426.91,266) .. (430.5,266) .. controls (434.09,266) and (437,268.91) .. (437,272.5) .. controls (437,276.09) and (434.09,279) .. (430.5,279) .. controls (426.91,279) and (424,276.09) .. (424,272.5) -- cycle ;
\draw  [draw opacity=0][fill={rgb, 255:red, 208; green, 2; blue, 27 }  ,fill opacity=1 ] (422,128.5) .. controls (422,124.91) and (424.91,122) .. (428.5,122) .. controls (432.09,122) and (435,124.91) .. (435,128.5) .. controls (435,132.09) and (432.09,135) .. (428.5,135) .. controls (424.91,135) and (422,132.09) .. (422,128.5) -- cycle ;

\end{tikzpicture}

%% file: sections/tikz/tikz-2dflows-2.tex
\tikzset{every picture/.style={line width=0.75pt}} 

\begin{tikzpicture}[x=0.75pt,y=0.75pt,yscale=-1,xscale=1]

\draw  [draw opacity=0][fill={rgb, 255:red, 80; green, 227; blue, 194 }  ,fill opacity=0.5 ][line width=1.5]  (67.68,86.25) -- (211.77,56.81) -- (216.32,79.08) -- (72.23,108.53) -- cycle ;
\draw  [draw opacity=0][fill={rgb, 255:red, 80; green, 227; blue, 194 }  ,fill opacity=0.5 ][line width=1.5]  (79.08,90.9) -- (141.53,174.53) -- (122.92,188.43) -- (60.47,104.8) -- cycle ;
\draw  [draw opacity=0][fill={rgb, 255:red, 80; green, 227; blue, 194 }  ,fill opacity=0.5 ][line width=1.5]  (35.97,30.07) -- (89.14,75.05) -- (50.03,121.27) -- (-3.14,76.29) -- cycle ;
\draw  [draw opacity=0][fill={rgb, 255:red, 80; green, 227; blue, 194 }  ,fill opacity=0.5 ][line width=1.5]  (131.58,170.63) -- (183.58,168.65) -- (184.42,190.7) -- (132.42,192.68) -- cycle ;
\draw  [draw opacity=0][fill={rgb, 255:red, 80; green, 227; blue, 194 }  ,fill opacity=0.5 ][line width=1.5]  (37.12,310.95) -- (297.24,340.86) -- (295.05,359.95) -- (34.92,330.04) -- cycle ;
\draw  [draw opacity=0][fill={rgb, 255:red, 80; green, 227; blue, 194 }  ,fill opacity=0.5 ][line width=1.5]  (199.99,145.24) -- (271.44,178.44) -- (240.01,246.09) -- (168.56,212.9) -- cycle ;
\draw  [draw opacity=0][fill={rgb, 255:red, 80; green, 227; blue, 194 }  ,fill opacity=0.5 ][line width=1.5]  (8.7,322.41) -- (30.24,310.41) -- (41.12,329.92) -- (19.57,341.93) -- cycle ;
\draw  [draw opacity=0][fill={rgb, 255:red, 80; green, 227; blue, 194 }  ,fill opacity=0.5 ][line width=1.5]  (11.38,304.65) -- (103.48,163.11) -- (160.01,199.89) -- (67.92,341.43) -- cycle ;
\draw  [draw opacity=0][fill={rgb, 255:red, 80; green, 227; blue, 194 }  ,fill opacity=0.5 ][line width=1.5]  (226.58,70.98) -- (197.02,183.1) -- (171.42,176.35) -- (200.98,64.23) -- cycle ;
\draw  [draw opacity=0][fill={rgb, 255:red, 80; green, 227; blue, 194 }  ,fill opacity=0.5 ][line width=1.5]  (337.13,303.42) -- (334.89,353.41) -- (256.87,349.91) -- (259.11,299.92) -- cycle ;
\draw  [draw opacity=0][fill={rgb, 255:red, 80; green, 227; blue, 194 }  ,fill opacity=0.5 ][line width=1.5]  (348.29,323.75) -- (408.98,437.46) -- (305.71,492.58) -- (245.02,378.87) -- cycle ;
\draw  [draw opacity=0][fill={rgb, 255:red, 80; green, 227; blue, 194 }  ,fill opacity=0.5 ][line width=1.5]  (267.28,206.72) -- (307.17,297.64) -- (286.72,306.62) -- (246.83,215.69) -- cycle ;
\draw  [draw opacity=0][fill={rgb, 255:red, 80; green, 227; blue, 194 }  ,fill opacity=0.5 ][line width=1.5]  (290.42,291.48) -- (423.9,192.13) -- (438.58,211.85) -- (305.1,311.2) -- cycle ;
\draw  [draw opacity=0][fill={rgb, 255:red, 80; green, 227; blue, 194 }  ,fill opacity=0.5 ][line width=1.5]  (232.81,195.71) -- (269.42,140.78) -- (317.19,172.63) -- (280.58,227.55) -- cycle ;
\draw  [draw opacity=0][fill={rgb, 255:red, 80; green, 227; blue, 194 }  ,fill opacity=0.5 ][line width=1.5]  (207.19,63.13) -- (223.5,38.67) -- (237.81,48.21) -- (221.5,72.67) -- cycle ;
\draw  [draw opacity=0][fill={rgb, 255:red, 80; green, 227; blue, 194 }  ,fill opacity=0.5 ][line width=1.5]  (287.86,147.47) -- (347.36,107.72) -- (360.14,126.86) -- (300.64,166.61) -- cycle ;
\draw  [draw opacity=0][fill={rgb, 255:red, 80; green, 227; blue, 194 }  ,fill opacity=0.5 ][line width=1.5]  (428.73,166.25) -- (490.67,162.29) -- (495.27,234.09) -- (433.33,238.05) -- cycle ;
\draw  [draw opacity=0][fill={rgb, 255:red, 80; green, 227; blue, 194 }  ,fill opacity=0.5 ][line width=1.5]  (381.71,91.34) -- (460.18,173.94) -- (403.29,227.99) -- (324.82,145.39) -- cycle ;
\draw  [draw opacity=0][fill={rgb, 255:red, 80; green, 227; blue, 194 }  ,fill opacity=0.5 ][line width=1.5]  (312.81,106.89) -- (341.67,-3.41) -- (425.19,18.44) -- (396.33,128.74) -- cycle ;
\draw  [draw opacity=0][fill={rgb, 255:red, 80; green, 227; blue, 194 }  ,fill opacity=0.5 ][line width=1.5]  (195.89,34.48) -- (204.79,0.46) -- (275.11,18.85) -- (266.21,52.88) -- cycle ;
\draw  [draw opacity=0][fill={rgb, 255:red, 80; green, 227; blue, 194 }  ,fill opacity=0.5 ][line width=1.5]  (240.21,38.54) -- (303.13,151.59) -- (287.42,160.33) -- (224.5,47.28) -- cycle ;

\draw  [draw opacity=0][fill={rgb, 255:red, 208; green, 2; blue, 27 }  ,fill opacity=1 ] (30,184.5) .. controls (30,180.91) and (32.91,178) .. (36.5,178) .. controls (40.09,178) and (43,180.91) .. (43,184.5) .. controls (43,188.09) and (40.09,191) .. (36.5,191) .. controls (32.91,191) and (30,188.09) .. (30,184.5) -- cycle ;
\draw [color={rgb, 255:red, 126; green, 211; blue, 33 }  ,draw opacity=1 ][line width=3]    (14,332) -- (36,320.67) ;
\draw [color={rgb, 255:red, 126; green, 211; blue, 33 }  ,draw opacity=1 ][line width=3]    (36,320.67) -- (296,351.67) ;
\draw [color={rgb, 255:red, 126; green, 211; blue, 33 }  ,draw opacity=1 ][line width=3]    (358,464.67) -- (296,351.67) ;
\draw [color={rgb, 255:red, 126; green, 211; blue, 33 }  ,draw opacity=1 ][line width=3]    (298,301.67) -- (296,351.67) ;
\draw [color={rgb, 255:red, 126; green, 211; blue, 33 }  ,draw opacity=1 ][line width=3]    (431,201.67) -- (298,301.67) ;
\draw [color={rgb, 255:red, 126; green, 211; blue, 33 }  ,draw opacity=1 ][line width=3]    (298,301.67) -- (256,211.67) ;
\draw [color={rgb, 255:red, 126; green, 211; blue, 33 }  ,draw opacity=1 ][line width=3]    (36,320.67) -- (132,181.67) ;
\draw [color={rgb, 255:red, 126; green, 211; blue, 33 }  ,draw opacity=1 ][line width=3]    (132,181.67) -- (184,179.67) ;
\draw [color={rgb, 255:red, 126; green, 211; blue, 33 }  ,draw opacity=1 ][line width=3]    (184,179.67) -- (256,211.67) ;
\draw [color={rgb, 255:red, 126; green, 211; blue, 33 }  ,draw opacity=1 ][line width=3]    (294,156.67) -- (256,211.67) ;
\draw [color={rgb, 255:red, 126; green, 211; blue, 33 }  ,draw opacity=1 ][line width=3]    (493,198.67) -- (431,201.67) ;
\draw [color={rgb, 255:red, 126; green, 211; blue, 33 }  ,draw opacity=1 ][line width=3]    (354,117.67) -- (431,201.67) ;
\draw [color={rgb, 255:red, 126; green, 211; blue, 33 }  ,draw opacity=1 ][line width=3]    (294,156.67) -- (354,117.67) ;
\draw [color={rgb, 255:red, 126; green, 211; blue, 33 }  ,draw opacity=1 ][line width=3]    (214,67.67) -- (184,179.67) ;
\draw [color={rgb, 255:red, 126; green, 211; blue, 33 }  ,draw opacity=1 ][line width=3]    (70,97.67) -- (132,181.67) ;
\draw [color={rgb, 255:red, 126; green, 211; blue, 33 }  ,draw opacity=1 ][line width=3]    (70,97.67) -- (214,67.67) ;
\draw [color={rgb, 255:red, 126; green, 211; blue, 33 }  ,draw opacity=1 ][line width=3]    (70,97.67) -- (16,53.67) ;
\draw [color={rgb, 255:red, 126; green, 211; blue, 33 }  ,draw opacity=1 ][line width=3]    (214,67.67) -- (231,43.67) ;
\draw [color={rgb, 255:red, 126; green, 211; blue, 33 }  ,draw opacity=1 ][line width=3]    (231,43.67) -- (294,156.67) ;
\draw [color={rgb, 255:red, 126; green, 211; blue, 33 }  ,draw opacity=1 ][line width=3]    (231,43.67) -- (240,9.67) ;
\draw [color={rgb, 255:red, 126; green, 211; blue, 33 }  ,draw opacity=1 ][line width=3]    (354,117.67) -- (384,7.67) ;
\draw  [draw opacity=0][fill={rgb, 255:red, 208; green, 2; blue, 27 }  ,fill opacity=1 ] (161,266.5) .. controls (161,262.91) and (163.91,260) .. (167.5,260) .. controls (171.09,260) and (174,262.91) .. (174,266.5) .. controls (174,270.09) and (171.09,273) .. (167.5,273) .. controls (163.91,273) and (161,270.09) .. (161,266.5) -- cycle ;
\draw  [draw opacity=0][fill={rgb, 255:red, 208; green, 2; blue, 27 }  ,fill opacity=1 ] (144,406.5) .. controls (144,402.91) and (146.91,400) .. (150.5,400) .. controls (154.09,400) and (157,402.91) .. (157,406.5) .. controls (157,410.09) and (154.09,413) .. (150.5,413) .. controls (146.91,413) and (144,410.09) .. (144,406.5) -- cycle ;
\draw  [draw opacity=0][fill={rgb, 255:red, 208; green, 2; blue, 27 }  ,fill opacity=1 ] (159,96.5) .. controls (159,92.91) and (161.91,90) .. (165.5,90) .. controls (169.09,90) and (172,92.91) .. (172,96.5) .. controls (172,100.09) and (169.09,103) .. (165.5,103) .. controls (161.91,103) and (159,100.09) .. (159,96.5) -- cycle ;
\draw  [draw opacity=0][fill={rgb, 255:red, 208; green, 2; blue, 27 }  ,fill opacity=1 ] (150,61.5) .. controls (150,57.91) and (152.91,55) .. (156.5,55) .. controls (160.09,55) and (163,57.91) .. (163,61.5) .. controls (163,65.09) and (160.09,68) .. (156.5,68) .. controls (152.91,68) and (150,65.09) .. (150,61.5) -- cycle ;
\draw  [draw opacity=0][fill={rgb, 255:red, 208; green, 2; blue, 27 }  ,fill opacity=1 ] (233,114.5) .. controls (233,110.91) and (235.91,108) .. (239.5,108) .. controls (243.09,108) and (246,110.91) .. (246,114.5) .. controls (246,118.09) and (243.09,121) .. (239.5,121) .. controls (235.91,121) and (233,118.09) .. (233,114.5) -- cycle ;
\draw  [draw opacity=0][fill={rgb, 255:red, 208; green, 2; blue, 27 }  ,fill opacity=1 ] (280,91.5) .. controls (280,87.91) and (282.91,85) .. (286.5,85) .. controls (290.09,85) and (293,87.91) .. (293,91.5) .. controls (293,95.09) and (290.09,98) .. (286.5,98) .. controls (282.91,98) and (280,95.09) .. (280,91.5) -- cycle ;
\draw  [draw opacity=0][fill={rgb, 255:red, 208; green, 2; blue, 27 }  ,fill opacity=1 ] (350,187.5) .. controls (350,183.91) and (352.91,181) .. (356.5,181) .. controls (360.09,181) and (363,183.91) .. (363,187.5) .. controls (363,191.09) and (360.09,194) .. (356.5,194) .. controls (352.91,194) and (350,191.09) .. (350,187.5) -- cycle ;
\draw  [draw opacity=0][fill={rgb, 255:red, 208; green, 2; blue, 27 }  ,fill opacity=1 ] (424,272.5) .. controls (424,268.91) and (426.91,266) .. (430.5,266) .. controls (434.09,266) and (437,268.91) .. (437,272.5) .. controls (437,276.09) and (434.09,279) .. (430.5,279) .. controls (426.91,279) and (424,276.09) .. (424,272.5) -- cycle ;
\draw  [draw opacity=0][fill={rgb, 255:red, 208; green, 2; blue, 27 }  ,fill opacity=1 ] (422,128.5) .. controls (422,124.91) and (424.91,122) .. (428.5,122) .. controls (432.09,122) and (435,124.91) .. (435,128.5) .. controls (435,132.09) and (432.09,135) .. (428.5,135) .. controls (424.91,135) and (422,132.09) .. (422,128.5) -- cycle ;

\end{tikzpicture}

%% file: sections/sec-numerical_flux.tex
\subsection{Numerical Flux} \label{sec:numerical_flux}

\input{sections/img/fig-methods-reflection_boundary}

As noted in Equation \ref{eq:godunov}, we require the evaluation of some numerical flux $\mathbf{\hat{F}}_{i,j}$ at each pairwise interface between neighbours $i$ and $j$.
Numerous methods exist to solve this, including an analytic solution via characteristic decomposition \cite{toro1994restoration}, and various linearization-based approximations \cite{roe1981approximate, harten1983upstream, einfeldt1988godunov}.
Our method is agnostic of this choice, and generally more accurate solvers exchange reduced diffusivity with increased computational cost.

In our implementation, we choose to take the numerical flux described in \citet{kurganov2000new}, which has been prior shown to be sufficiently diffusive to reduce noise that would lead to multivalued fields. They use:
\begin{align}
    \hat{\mathbf{F}}_{ij} = \frac{1}{2}\left(\mathbf{F}_{ij}^{-}+\mathbf{F}_{ij}^{+}\right) - \frac{a_{ij}}{2}\left(\mathbf{U}_{ij}^{+}-\mathbf{U}_{ij}^{-}\right). \label{eq:kt2000}
\end{align}
To define the signal velocity $a_{ij}$, consider a path $\mathcal{C}\left(\mathbf{U}_{ij}^-,\mathbf{U}_{ij}^+\right)$, connecting the left and right states of a Riemann fan.
The signal velocity is the spectral radius of the flux Jacobian $\partial\mathbf{F}/\partial\mathbf{U}$ evaluated along this curve.
In other words, it is the largest magnitude eigenvalue along the path:
\begin{align}
    a_{ij} &= \max_{\mathbf{U}\in\mathcal{C}\left(\mathbf{U}_{ij}^-,\mathbf{U}_{ij}^+\right)}\lambda(\mathbf{U}).
\end{align}
For the compressible Euler equations, the eigenvalues of the flux Jacobian are $\lambda(\mathbf{U}) = \left\{u_n, u_n+c, u_n-c\right\}$.
For a 1D Riemann problem, this is simply evaluates to the maximum eigenvalue at either side of the interface, $\max\left\{|u_n^-\pm c^-|, |u_n^+\pm c^+|\right\}$, with $c=\sqrt{\gamma P/\rho}$ being the sound speed of the fluid.

Conceptually, these numerical fluxes form waves traveling out from each interface, moving at the speed $a_{ij}$, as sketched out on Figure \ref{fig:methods-voronoi-flows}.
Notice then that $a_{ij}$ exactly determines the timestep restriction--waves cannot be allowed to travel past the middle of the cell.
Outside this wave propagation domains, the solution remains the same from the previous timestep.

%% file: sections/img/fig-methods-reflection_boundary.tex
\begin{figure}[t]
  \centering
  \includegraphics[width=0.8\linewidth]{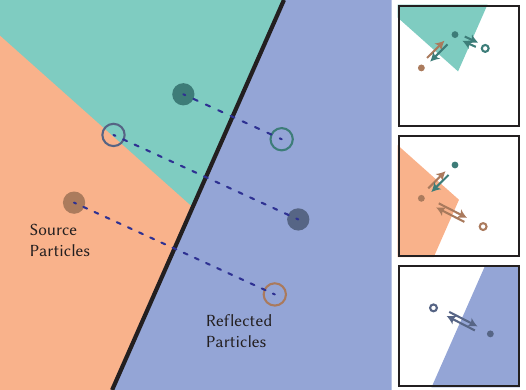}
  \caption{Boundary condition enforcement via reflected particles. Reflected particles are depicted as outlined circles in the global state above, but only exist in the context of their corresponding particles in order to resolve the solid boundary condition (right). Note that locally, the solid interface is resolved exactly the same way as fluid interfaces, with a flux computed along the face.} \label{fig:methods-reflection_boundary}
  \Description{}
\end{figure}

%% file: sections/sec-boundary_condition.tex
\subsection{Boundary Enforcement} \label{sec:boundary_enforcement}

\input{sections/img/fig-r-bunny}

\input{sections/img/fig-r-bunny_velocity}

Because our discretization includes the solid boundaries as faces in the fluid partition, we greatly simplify the approach to respecting solid boundaries.
We simply must enforce zero flux across the interface in the frame of the solid face's velocity, thus dictating that the fluid must match the solid face velocity at the interface.

To do so, for every solid face in a particle's fluid cell boundary, we create a reflected particle across this solid face, as shown on Figure \ref{fig:methods-reflection_boundary}.
These reflected particles are entirely local, and exist only from the point of view of the particle being reflected.
They do not exist in the actual simulation domain, and do not participate in any other particle's flux computation.
This differs from approaches with explicit solid Voronoi cells (\cite{springel2010pur, hwang2021coupling}), which necessarily introduces a volume to the solid because of its inclusion in the Voronoi.
We additionally note that this also allows us to resolve even subgrid boundaries immersed entirely within a cell, as those faces are still added to the mesh and enforce this fluid-solid boundary flux.

These reflected particles share the original particle's density and pressure, and will have a velocity mirrored in the face normal direction in the solid velocity's frame.
Taking together the boost to solid velocity frame, performing the reflection, and deboosting to the lab frame, we have,
\begin{align}
    \mathbf{u}_{f} = \mathbf{u}_f - 2((\mathbf{u}_p-\mathbf{u}_s)\cdot \mathbf{n})\mathbf{n},
\end{align}
with $\mathbf{u}_f$ being the fluid particle velocity, $\mathbf{u}_s$ being the solid face velocity, and $\mathbf{n}$ is the outwards-facing normal.

%% file: sections/img/fig-r-bunny.tex
\begin{figure}[t]
  \centering
  \includegraphics[width=\linewidth]{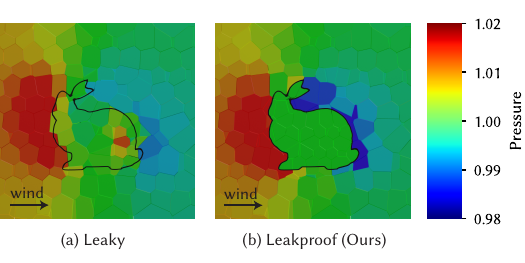}
  \caption{Cutaway view of a bunny inside a windtunnel without (a) and with (b) our Voronoi stitching method. The fluid inside the watertight bunny remains quiescent using our method, but gains velocity via leaking through the interface using a naive approach.}
  \label{fig:r-bunny}
\end{figure}

%% file: sections/img/fig-r-bunny_velocity.tex
\begin{figure}[b]
  \centering
  \includegraphics[width=\linewidth]{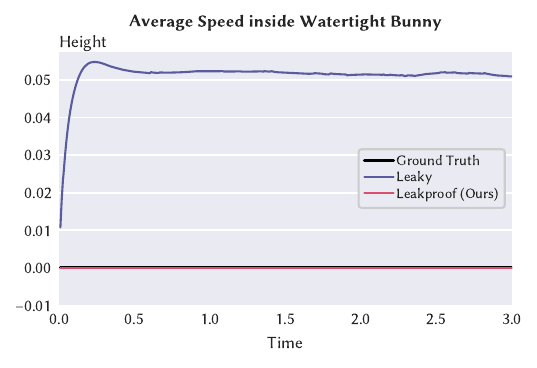}
  \caption{Average speed over time inside a watertight bunny inside a windtunnel. Our leakproof partitioning keeps the interior quiescent for the duration of the simulation, matching the ground truth, while the naive approach gains 50\% of the exterior velocity.} \label{fig:r-bunny_velocity}
  \Description{}
\end{figure}

%% file: sections/sec-examples.tex
\section{Results}

\input{sections/tables/tbl-scene_params}

We implemented our method as a solver node in Houdini \cite{houdini}, using its  built in rigid body and FEM solvers for solid simulation \cite{houdini}.
Scene parameters are provided in Table \ref{tbl:scene_params}.

\input{sections/img/fig-r-airspring_visc_noise}

\subsection{Sealed Bunny}

We place a static bunny boundary inside a wind tunnel, shown on Figure \ref{fig:r-bunny}.
The exterior fluid is initialized with a velocity of $u_x=0.1$ lengthwise across the bunny, while the interior fluid is initialized with zero velocity.
Inflow and outflow boundaries are placed at either end of the wind tunnel to maintain the exterior fluid velocity.

We simulate this watertight bunny boundary with both our leak-proof Voronoi stitching algorithm, and a naive algorithm of just using the existing Voronoi mesh with immersed solid boundary treatment.
That is, the immersed solids still enforce the defined in Section \ref{sec:boundary_enforcement}.
We find that our leakproof treatment retains the original fluid values up to machine single-precision, while the leaky bunny quickly gains velocity induced by the exterior wind, as shown on Figure \ref{fig:r-bunny_velocity}.

As a stress test of our stitching algorithm, we also simulated this with the same exterior fluid particles, but seed only a single fluid particle on the bunny interior.
We find that the stitching algorithm is able to correctly assign all cells inside the bunny to that single particle.
This means the entire geometry is represented by a single particle's state.
This state remained fully watertight and experienced no flux with the external fluid, thus retaining its original values up to machine precision.

\subsection{Air Spring}

\input{sections/img/fig-r-airspring_renders}

Figure \ref{fig:r-airspring_visc_noise} shows the displacement from a simulated air spring, consisting of a sealed vertical chamber with two pistons constrained to vertical motion.
The bottom piston is a scripted driver piston, while the top piston is allowed to freely move while experiencing gravity pulling it down and the pressure of the enclosed air pushing it up.

Demonstrating that our method can be expanded into viscous fluids, we introduce an extra viscous flux term to our numerical flux:
\begin{align}
    F^{visc}\cdot\mathbf{n}=\begin{bmatrix}
        0 \\
        \tau_{xj}n_j \\
        \tau_{yj}n_j \\
        \tau_{zj}n_j \\
        u_i\tau_{ij}n_j - q_jn_j
    \end{bmatrix},
\end{align}
where $i,j$ are Einstein summation indices for three spatial dimensions. 
$\tau_{ij}=\mu\left(\frac{\partial u_i}{\partial x_j} + \frac{\partial u_j}{\partial x_i} - \frac{2}{3}\delta_{ij}\nabla\cdot \mathbf{u}\right)$ produces the viscous stress tensor with dynamic viscosity $\mu$, and the heat flux is given by $q_j=-\kappa\frac{\partial T}{\partial x_j}$ with thermal conductivity $\kappa$.
Temperature $T$ was computed via ideal gas law $P/\rho=RT$, with $R$ taken to be $1$ in simulation units for the purpose of this example.

A render of a single travelling pulse is shown on Figure \ref{fig:r-airspring_render}.
We see that since sound speed decreases as density increases ($c_s=\sqrt{\gamma P/\rho}$), the follower piston takes longer to respond with increasing density.
This has the added benefit of stretching out waves, thus smoothing out and reducing the follower's displacement.
We also see that increasing viscosity has a similar effect, blending out sharp waves caused by sudden movements of the driver.

\subsection{Stomp Rocket}

A rocket mass is placed above an air-filled tube, connected to an air-filled reservoir through a u-bend. 
A driver ``foot'' is rapidly pushed down, sending air through the u-bend and into the reservoir, where it pushes the rocket out; we plot height of the rocket over time on Figure \ref{fig:r-stomprocket}.
We test a variety of sound speed by modulating the density of the internal fluid, and see that fluids with lower sound speeds achieve higher overall heights.

Although the trajectory of the foot is the same for all tests, the work done is larger for lower sound speeds.
As sound speed decreases, the propagating wave moves slower, therefore more gas ``piles'' up, increasing pressure and therefore total work done.
This increased work translates to faster upwards velocity on the rocket.
On the opposite limit is the incompressible case, where the rocket will only travel the same distance as the foot travels down.

\input{sections/img/fig-r-stomprocket_stacked}

\subsection{Asteroid} \label{sec:asteroid}

To demonstrate that our method handles both subsonic and supersonic flow around solids, we simulate fluid flow around Asteroid 4486 Mithra \cite{nasa_mithra_2025}
Three different flow speeds and their corresponding mach numbers are shown on Figure \ref{fig:r-asteroid}, with the ground truth Mach cones shown as white lines.

\input{sections/img/fig-r-asteroid}

We note one deficiency with our current method is the loss of resolution in the supersonic wake.
Because the fluid particles move with the velocity, the wakes tend to become underresolved.
This is especially problematic for highly supersonic flow, such as the rightmost case, where no particles at all are found in the wake.
Our Voronoi discretization fills this in with the high pressure data from the neighbouring particles, improperly representing the low pressure wake.
This may be resolved with careful particle splitting whenever particle density becomes too low.

\subsection{Champagne Cork}

We simulate the ejection of a champagne cork from a bottle, where internal pressure gradually increases, representing dissolved $CO_2$ bubbling out of the solution.
The cork is initially held in place by friction between the cork and the neck, which is overcome by the increasing pressure of the air pocket, after which the cork is released and shot out of the bottle.
Our method seamlessly handles the topology change as the interior domain merges with the exterior domain, allowing fluid to start flowing outwards.

\input{sections/img/fig-r-winebottle}

\subsection{Balloon Propulsion}

We allow an inflated balloon to release its stored air, propelling itself through the simulation domain, as shown on Figure \ref{fig:teaser}. This demonstrates complex energy transfer between stored potential energy of the elastic balloon, potential energy of high pressure air, into fluid kinetic energy as air escapes the balloon due to the pressure difference inside and outside the balloon, to kinetic energy of the balloon as flies around the domain.

\subsection{Fan}

\input{sections/img/fig-r-fan}

We push air using a scripted rotating fan along two different duct geometries, shown on Figure \ref{fig:r-fan}.
We note that this example relies heavily on the resolution of subgrid surfaces, as the fan is considerably thinner than the particle resolution.
Even here, our method is able to transfer velocity from the solid to the fluid phase.

We see that the circular duct is able to more efficiently accelerate air, as shown by the greater pressure gradient across the fan, with the square duct demonstrating a considerable amount of backflow at the corners, shown by the negative axial velocities. 

%% file: sections/tables/tbl-scene_params.tex
\begin{table*}[t] 
\caption{Scene parameters for presented examples. $\mathbf{U}_L$ and $\mathbf{U}_R$ represent the left and right states (or exterior and interior states, as prescribed in the problem) respectively, with $\rho$, ${u}$, $P$ being fluid density, velocity, and pressure. Velocity is given as a scalar for the axis relevant to the problem, all other velocity components are $0$. $\#P$ is the number of fluid particles and $\#V$ is the number of solid vertices. 
Machines used for computation were: an M1 Macbook Pro with 32GB RAM (Mac), an 8-core AMD R7 1700 CPU with 64GB RAM (PC), and a 32-core AMD Threadripper 2990WX with 128 GB of RAM (Server).} \label{tbl:scene_params}
\begin{tabular}{lllllllllllp{2cm}ll}
\hline
\textit{}                                                 & \multicolumn{3}{l}{\textit{$\mathbf{U}_L$}}                                                                                 & \multicolumn{3}{l}{\textit{$\mathbf{U}_R$}}                                                                           & \multicolumn{3}{l}{\textit{}}                                                                                               &                \textit{Fluid}                 & \textit{Solid}          & \textit{}        & \multicolumn{1}{l}{\textit{}}                \\
\textit{}                                                 & \multicolumn{1}{l}{\textit{$\rho$}}                  & \textit{$u$} & \multicolumn{1}{l}{\textit{$P$}}             & \textit{$\rho$}                               & \textit{$u$} & \textit{$P$}                                  & \textit{Domain} & \multicolumn{1}{l}{\textit{$\#$Frames}}      & \multicolumn{1}{l}{\textit{Time}}            & \multicolumn{1}{l}{\textit{$\#P$}}           & \textit{$\#V$}                        & \textit{Machine} & \multicolumn{1}{l}{\textit{s/Frame}}         \\
\rowcolor[HTML]{C6D9EB} 
\textbf{Sealed Bunny}                                              & \multicolumn{1}{l}{\cellcolor[HTML]{C6D9EB}}         &                       & \multicolumn{1}{l}{\cellcolor[HTML]{C6D9EB}} &                                               &                       &                                               &                 & \multicolumn{1}{l}{\cellcolor[HTML]{C6D9EB}} & \multicolumn{1}{l}{\cellcolor[HTML]{C6D9EB}} & \multicolumn{1}{l}{\cellcolor[HTML]{C6D9EB}} &                                       &                  & \multicolumn{1}{l}{\cellcolor[HTML]{C6D9EB}} \\
\rowcolor[HTML]{E4EDF5} 
\multicolumn{1}{r}{\cellcolor[HTML]{E4EDF5}Leakproof}     & 1                                                    & 0.1          & 1                                            & \multicolumn{1}{r}{\cellcolor[HTML]{E4EDF5}1} & 0             & \multicolumn{1}{r}{\cellcolor[HTML]{E4EDF5}1} & 2 x 2 x 2       & 360                                          & 3                                            & 2652                                         & 6938 Static                           & Mac              & 29.67                                        \\
\rowcolor[HTML]{C6D9EB} 
\multicolumn{1}{r}{\cellcolor[HTML]{C6D9EB}Leaky}         & 1                                                    & 0.1          & 1                                            & \multicolumn{1}{r}{\cellcolor[HTML]{C6D9EB}1} & 0             & \multicolumn{1}{r}{\cellcolor[HTML]{C6D9EB}1} & 2 x 2 x 2       & 360                                          & 3                                            & 2652                                         & 6948 Static                           & Mac              & 24.00                                        \\
\textbf{Air Spring}                                                & \multicolumn{1}{l}{Var.}                         & 0             & 4                                            & -                                             & -                     & -                                             & 1 x 9 x 1       & 7200                                         & 15                                           & 672                                          & 263 Scripted\newline 263 Rigid\newline 433 Static & Server           & 1.74                                         \\
\rowcolor[HTML]{C6D9EB} 
\textbf{Stomp Rocket}                                              & \multicolumn{1}{l}{\cellcolor[HTML]{C6D9EB}Var.} & 0             & 4                                            & -                                             & -                     & -                                             & 3 x 4 x 2       & 1200                                         & 5                                            & 955                                          & 262 Scripted\newline 258 Rigid\newline 816 Static & Server           & 0.85                                         \\
\textbf{Asteroid}                                                  & 1                                                    & Var.              & 1                                            & -                                             & -                     & -                                             & 1 x 1 x 1       & 1920                                         & 2                                            & 22336                                        & 674 Rigid                             & Server           & 135.91                                       \\
\rowcolor[HTML]{C6D9EB} 
\textbf{Champagne}                                                 & 1                                                    & 0             & 1                                            & 10                                      & 0             & 10                                      & 1 x 1 x 1       & 1920                                         & 1                                            & 732                                          & 463 Rigid\newline 8382 Static               & PC               & 30.94                                        \\
\textbf{Balloon}                                                   & 1                                                    & 0             & 6                                            & -                                             & -                     & -                                             & 10 x 10 x 10    & 4800                                         & 10                                           & 1350                                         & 824~Deformable                        & PC               & 8.25                                         \\
\rowcolor[HTML]{C6D9EB} 
\textbf{Fan}                                                       & \multicolumn{1}{l}{\cellcolor[HTML]{C6D9EB}}         &                       & \multicolumn{1}{l}{\cellcolor[HTML]{C6D9EB}} &                                               &                       &                                               &                 & \multicolumn{1}{l}{\cellcolor[HTML]{C6D9EB}} & \multicolumn{1}{l}{\cellcolor[HTML]{C6D9EB}} & \multicolumn{1}{l}{\cellcolor[HTML]{C6D9EB}} &                                       &                  & \multicolumn{1}{l}{\cellcolor[HTML]{C6D9EB}} \\
\rowcolor[HTML]{E4EDF5} 
\multicolumn{1}{r}{\cellcolor[HTML]{E4EDF5}Square Duct}   & 1                                                    & 0             & 20                                           & -                                             & -                     & -                                             & 1.5 x 1.5 x 5   & 4800                                         & 20                                           & 5512                                         & 2395 Scripted                         & Server           & 16.32                                        \\
\rowcolor[HTML]{C6D9EB} 
\multicolumn{1}{r}{\cellcolor[HTML]{C6D9EB}Circular Duct} & 1                                                    & 0             & 20                                           & -                                             & -                     & -                                             & 1.5 x 1.5 x 5   & 4800                                         & 20                                           & 3703                                         & 2395 Scripted\newline 2391 Static           & Server           & 16.25                                        \\ \hline
\end{tabular}
\end{table*}


%% file: sections/img/fig-r-airspring_visc_noise.tex
\begin{figure}[t]
  \centering
  \includegraphics[width=\linewidth]{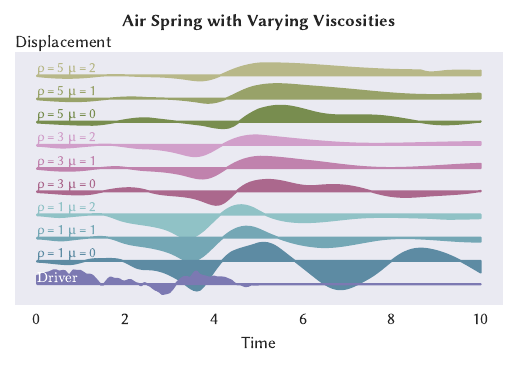}
  \caption{Displacement over time in an airspring with a scripted driver input. Air densities and dynamic viscosities are as marked, with thermal conductivity taken to be $\kappa=3\mu$.} \label{fig:r-airspring_visc_noise}
  \Description{}
\end{figure}

%% file: sections/img/fig-r-airspring_renders.tex
\begin{figure}[b]
  \centering
  \includegraphics[width=\linewidth]{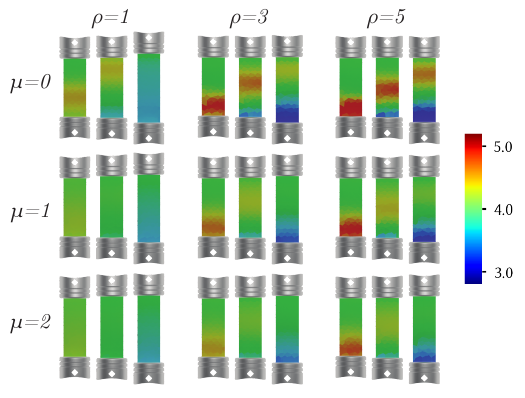}
  \caption{Cutaway renders of pressure for airsprings of various densities and dynamic viscosities depicting a single propagating sine wave at $t=0.4,0.8,1.2$.} \label{fig:r-airspring_render}
  \Description{}
\end{figure}

%% file: sections/img/fig-r-stomprocket_stacked.tex
\begin{figure}[b]
  \centering
  \includegraphics[width=\linewidth]{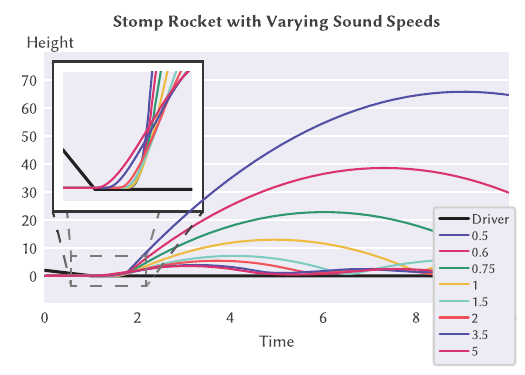}
  \caption{Displacement over time of a stomp rocket with a zoom in of early time inset.}
  \label{fig:r-stomprocket}
\end{figure}

%% file: sections/img/fig-r-asteroid.tex
\begin{figure}[t]
  \centering
  \includegraphics[width=\linewidth]{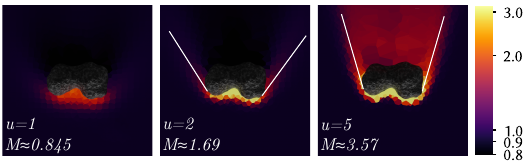}
  \caption{Cutaway view of pressure around an asteroid falling at various speeds, with the corresponding Mach number as marked. Ground truth Mach cones are shown as white lines for the two supersonic cases.}
  \label{fig:r-asteroid}
\end{figure}

%% file: sections/img/fig-r-winebottle.tex
\begin{figure}[t]
  \centering
  \includegraphics[width=\linewidth]{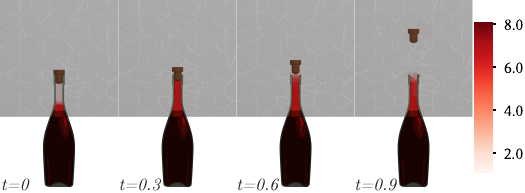}
  \caption{Cutaway view of pressure in a champagne bottle launching a cork held on by friction. Pressure inside the bottle is initially at 2 atm, and slowly rises to 10 atm via an inflow boundary at the wine surface. Outer domain consists of outflow boundary conditions.}
  \label{fig:r-winebottle}
\end{figure}

%% file: sections/img/fig-r-fan.tex
\begin{figure}[t]
  \centering
  \includegraphics[width=\linewidth]{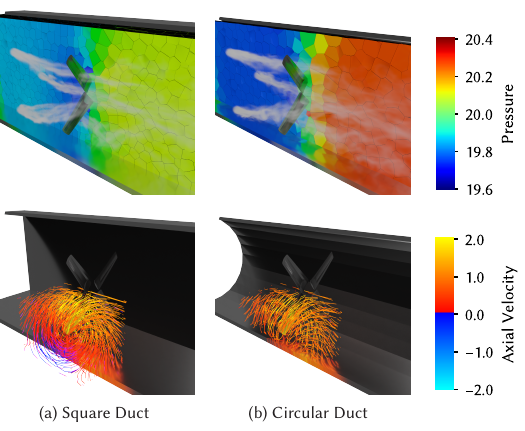}
  \caption{Cutaway view of the pressure gradient induced as a rotating fan pushes air forwards (above), and streamlines coloured by the axial flow velocity (below), for two different duct geometries. Smoke was generated in post-processing, using our computed velocity field.} \label{fig:r-fan}
\end{figure}

%% file: sections/sec-conclusion.tex
\section{Discussion and Conclusions}

We presented a method for coupling compressible fluids with solids while respecting the connectivity of the fluid, creating a discretization that is leakproof only when required, and flow permissive otherwise.
Our method leverages the Voronoi diagram to partition space, modifying it to conform to solid boundaries and allow for natural handling of force transfer between the solid and fluid regimes.

We identify extensions of our work that are ripe avenues for future research.
In particular, because of the explicit nature of our method, flux computation and advection are extremely fast and embarrassingly parallelizable.
Our method, however, is significantly bottlenecked by the cost of computing the Voronoi diagram.
Our current implementation uses Houdini's built-in library, which recomputes the Voronoi structure at each timestep, taking up $85\%$ of our simulation runtime.
Leveraging temporal coherence to update the previous timestep's structure is a promising method for accelerating \cite{guibas2018kinetic}.
We, however, note that the diagram is only required for connectivity and face areas.
If there was a method for approximating face areas while maintaining the same connectivity properties without fully computing a Voronoi diagram, the method may become significantly faster.

Additionally, the modifications to the Voronoi does sidestep the true intended structure, which is a visibility-constrained Voronoi.
That is, every cell should have the domain of every point in space closest to it, while accounting for visibility constraints (the solid boundaries).
Imagine the source points growing in the domain given some occluding planes, the points would propagate out waves that naturally bend around the ends of the occluding planes.
Some work exists for rectilinear occluders in 2D \cite{tsin1996geodesic}, but extensions to arbitrary barriers in 3D remain an open problem.
Similar work can also be used for more accurate intracell gradients, as gradients to faces should vary according to occluders within these cells.

We also found that in certain examples, such as the supersonic wake in Section \ref{sec:asteroid}, the Lagrangian particles become underresolved.
This could be alleviated via a particle splitting scheme wherever local particle density becomes too low.
We emphasize, however, that our method still maintains our desired connectivity property even when resolution drops like this, with the neighbouring particles simply extending their domains into the underresolved regions.

Finally, because of our choice of explicit method, coupling is handled by the multiple timestepping.
That is, fluid imposes some boundary condition on the solid, which evolves and imposes a boundary condition onto the liquid.
A more accurate approach would be to find the force balance between the two regimes \emph{within} each timestep, such as with a Newton iterative solver.

Research on compressible flow is typically driven by applications in science and engineering, some of which fall under ethically grey domains.
Beyond technical novelty, we hope that our work highlights the rich visual and physical phenomena that emerge from advancing algorithms in comppressible flow, and that it encourages further exploration in creative and constructive directions.